\documentclass[aps,pra,twocolumn,reprint]{revtex4-2}

\usepackage{graphicx}
\usepackage{amsmath, amssymb, amsfonts}
\usepackage{bm}
\usepackage{xcolor}
\usepackage[normalem]{ulem}

\DeclareMathAlphabet\mathbfcal{OMS}{cmsy}{b}{n}

\newlength{\figwidth}
\newlength{\lfig}
\newlength{\sfig}
\begin{document}
\title{Making molecules by mergoassociation: \\ two atoms in adjacent nonspherical optical traps}

\author{Robert C. Bird}
\affiliation{Joint Quantum Centre (JQC) Durham-Newcastle, Department of
Chemistry, Durham University, South Road, Durham, DH1 3LE, United Kingdom.}
\author{C. Ruth Le Sueur}
\affiliation{Joint Quantum Centre (JQC) Durham-Newcastle, Department of
Chemistry, Durham University, South Road, Durham, DH1 3LE, United Kingdom.}
\author{Jeremy M. Hutson}
\email{j.m.hutson@durham.ac.uk} \affiliation{Joint Quantum Centre (JQC)
Durham-Newcastle, Department of Chemistry, Durham University, South Road,
Durham, DH1 3LE, United Kingdom.}

\date{\today}

\begin{abstract}
Mergoassociation of two ultracold atoms to form a weakly bound molecule can occur when two optical traps that each contain a single atom are merged. Molecule formation occurs at an avoided crossing between a molecular state and the lowest motional state of the atom pair. We develop the theory of mergoassociation for pairs of nonidentical nonspherical traps. We develop a coupled-channel approach for the relative motion of the two atoms and present results for pairs of cylindrically symmetrical traps as a function of their anisotropy. We focus on the strength of the avoided crossing responsible for mergoassociation. We also develop an approximate method that gives insight into the dependence of the crossing strength on aspect ratio.
\end{abstract}

\maketitle

\section{Introduction}

Ultracold molecules have recently been formed in optical tweezers by mergoassociation \cite{Ruttley:2023}. The process begins with two atoms in separate tweezer traps, which are then merged. The atom pair is converted into a molecule by the merging process, with no further action required.

\begin{figure}[tbp]
\begin{center}
\includegraphics[width=0.8\columnwidth]{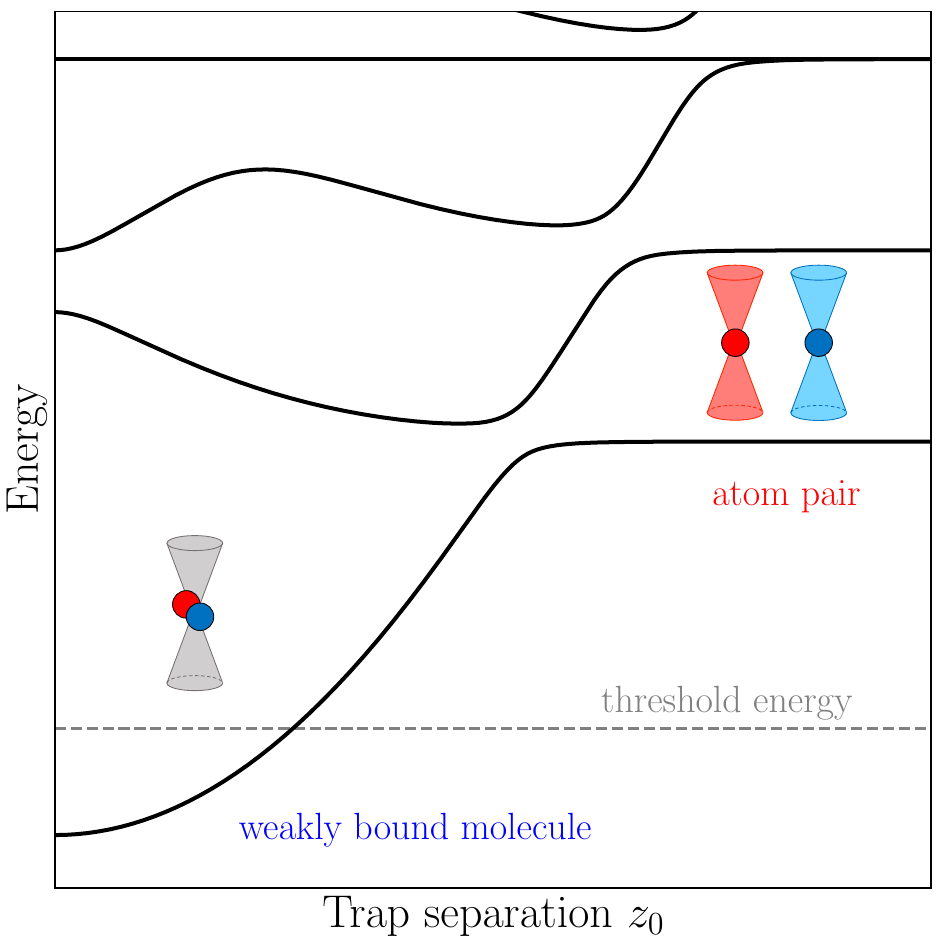}
	\caption{Schematic representation of the energy levels involved in mergoassociation, as a function of trap separation $z_0$. The molecular level (approximately quadratic as a function of $z_0$) has avoided crossings with motional states of the atom pair (approximately horizontal at large $z_0$). Mergoassociation occurs when an atom pair in the lowest motional state is transferred into the molecular state by adiabatic passage over the lowest avoided crossing.
\label{fig:intro}
}
\end{center}
\end{figure}

The energy levels involved in mergoassociation are shown schematically in Fig.\ \ref{fig:intro}. As a function of trap separation, there is an avoided crossing between the lowest motional state of the atom pair and a weakly bound molecular state. If the merging is carried out slowly enough to follow the crossing adiabatically, the atom pair is converted into a weakly bound molecule. A major advantage of this approach is that it can work even for unstructured atoms, and does not require a magnetically tunable Feshbach resonance. It thus opens the way to creating ultracold molecules from atom pairs without Feshbach resonances at experimentally accessible magnetic fields. It also offers possibilities for constructing two-qubit gates for quantum logic operations \cite{Jaksch:1999}.

Avoided crossings between atomic and molecular states as a function of trap separation were first studied by Stock \emph{et al.}\ \cite{Stock:2003, Stock:2006}, who considered the case of two atoms initially in identical spherical traps. Other authors have investigated similar situations for ions and molecules in spherical or quasi-1d traps \cite{Idziaszek:2007, Sroczynska:2022}. However, optical tweezers for ultracold atoms are usually formed in the high-intensity region at the waist of a focused laser beam \cite{Kaufman:2012}. Such tweezers are strongly anisotropic, usually with much weaker confinement along the laser beam than perpendicular to it. In ref.\ \cite{Ruttley:2023}, the ratio of the corresponding harmonic frequencies was about 1:6. Nevertheless, simulating the experiment with the tweezers approximated as spherical gave surprisingly good agreement between experiment and theory.

In the present paper, we develop the theory of merging nonidentical nonspherical traps. In Section \ref{sec:separation} we derive the separation between the relative and center-of-mass motions for separated traps, including the coupling term between them. In Section \ref{sec:coupled-form}, we develop a numerically exact coupled-channel approach to handle the relative motion of two atoms in nonspherical traps, including the case of traps that are not coaligned. In Section \ref{sec:coupled-res}, we solve the coupled equations and present energy-level diagrams for merging of two cylindrically symmetric tweezers as a function of their aspect ratios. We focus on the strength of the lowest avoided crossing, which is the key quantity for mergoassociation, and show that it depends strongly on aspect ratio. The results nevertheless explain the success of the spherical approximation in {ref.}~\cite{Ruttley:2023}. In Section \ref{sec:approx} we develop an approximate method based on a basis-set approach, which qualitatively reproduces the coupled-channel results and gives insight into the dependence of avoided-crossing strength on aspect ratio. Finally, in Section \ref{sec:conc} we present conclusions and perspectives for future work.

\section{Separation of relative and center-of-mass motion}
\label{sec:separation}

We consider two atoms independently confined in adjacent optical traps. Atom $i$ has mass $m_i$ and position $\boldsymbol{R}_i$ and is confined in a trap centered at $\boldsymbol{R}_i^0$. The motion may be factorized approximately into terms involving the relative and center-of-mass coordinates of the pair, $\boldsymbol{R}$ and $\mathbfcal{R}$ respectively. The 2-atom kinetic energy operator is exactly separable,
\begin{equation}
-\frac{\hbar^2}{2m_1}\nabla_1^2 - \frac{\hbar^2}{2m_2}\nabla_2^2 = -\frac{\hbar^2}{2{\cal M}}\nabla_{\cal R}^2 - \frac{\hbar^2}{2\mu}\nabla_R^2,
\end{equation}
where
\begin{align}
\mathbfcal{R}   &= \left(m_1 \boldsymbol{R}_1   + m_2 \boldsymbol{R}_2  \right)/{\cal M};\\
\boldsymbol{R}   &= \boldsymbol{R}_2   - \boldsymbol{R}_1;\\
{\cal M} &= m_1+m_2;\\
\mu &= m_1 m_2 / {\cal M}.
\end{align}

\subsection{Spherical traps}
\label{sec:spher}

If each trap is harmonic and spherical, the total potential energy due to the traps is
\begin{equation}
V^\textrm{trap}  = \textstyle{\frac{1}{2}} m_1 \omega_1^2 |\boldsymbol{R}_1-\boldsymbol{R}_1^0|^2
+ \textstyle{\frac{1}{2}} m_2 \omega_2^2 |\boldsymbol{R}_2-\boldsymbol{R}_2^0|^2,
\end{equation}
where $\omega_i$ is the harmonic frequency for atom $i$.
This may be written
\begin{align}
\textstyle{\frac{1}{2}} \mu \omega_\textrm{rel}^2 |\boldsymbol{R}-\boldsymbol{R}_0|^2
+&\textstyle{\frac{1}{2}} {\cal M} \omega_\textrm{com}^2 |\mathbfcal{R}-\mathbfcal{R}_0|^2 \nonumber\\
+& \mu \Delta\omega^2 (\mathbfcal{R}-\mathbfcal{R}_0) \cdot (\boldsymbol{R}-\boldsymbol{R}_0),
\label{eq:sep}
\end{align}
where
\begin{align}
\boldsymbol{R}_0 &= \boldsymbol{R}_2^0 - \boldsymbol{R}_1^0;\\
\omega_\textrm{rel}^2 &= \left({m_2\omega_1^2+m_1\omega_2^2}\right)\big/{\cal M};\\
\mathbfcal{R}_0 &= \left(m_1 \boldsymbol{R}_1^0 + m_2 \boldsymbol{R}_2^0\right)/{\cal M};\\
\omega_\textrm{com}^2 &= \left(m_1\omega_1^2+m_2\omega_2^2\right)/{\cal M};\\
\Delta\omega^2 &= \omega_2^2 - \omega_1^2.
\label{eq:conf-freq}
\end{align}
This is a generalization of the result of Stock \emph{et al.}\ \cite{Stock:2003}, who dealt with the case $m_1=m_2$ and $\omega_1=\omega_2$, so that the coupling term vanished.
The separation is similar to that for two nonidentical atoms in a single trap \cite{Deuretzbacher:2008}, except that the coupling term here involves $(\mathbfcal{R}-\mathbfcal{R}_0) \cdot (\boldsymbol{R}-\boldsymbol{R}_0)$ instead of $\mathbfcal{R} \cdot \boldsymbol{R}$. The relative and center-of-mass motions are uncoupled if the trap frequencies for the two atoms are the same. The coupling is generally not important if both atoms are in the motional ground state, but can be significant when trap states that are excited in the relative and center-of-mass motions are nearly degenerate.

\subsection{Nonspherical traps}
\label{sec:nonspher}

If the individual traps are harmonic but non-spherical, each trap has three principal axes perpendicular to one another. Eq.\ \ref{eq:sep} generalizes to
\begin{align}
V^\textrm{trap}&=\textstyle{\frac{1}{2}} \mu [\boldsymbol{R}-\boldsymbol{R}_0]^T \boldsymbol{\omega}_\textrm{rel}^2 [\boldsymbol{R}-\boldsymbol{R}_0] \nonumber\\
&+\textstyle{\frac{1}{2}} {\cal M} [\mathbfcal{R}-\mathbfcal{R}_0]^T \boldsymbol{\omega}_\textrm{com}^2 [\mathbfcal{R}-\mathbfcal{R}_0] \nonumber\\
&+ \mu [\mathbfcal{R}-\mathbfcal{R}_0]^T \Delta\boldsymbol{\omega}^2 [\boldsymbol{R}-\boldsymbol{R}_0],
\label{eq:sep-nonspher}
\end{align}
where $\boldsymbol{\omega}_\textrm{rel}^2$, $\boldsymbol{\omega}_\textrm{com}^2$ and $\Delta\boldsymbol{\omega}^2$ are second-rank tensors.
We choose Cartesian axes $X,Y,Z$ along the principal axes of
$\boldsymbol{\omega}_\textrm{rel}^2$, so that it may be represented as a diagonal matrix,
\begin{equation}
\boldsymbol{\omega}_\textrm{rel}^2=\begin{pmatrix}\omega_{{\rm rel},x}^2 & 0 & 0 \\ 0 & \omega_{{\rm rel},y}^2 & 0 \\ 0 & 0 & \omega_{{\rm rel},z}^2\end{pmatrix}.
\end{equation}
If the two traps are coaligned, meaning that they share the same set of principal axes, $\boldsymbol{\omega}_\textrm{com}^2$ and $\Delta\boldsymbol{\omega}^2$ are also diagonal matrices, defined similarly; if not, they are nondiagonal symmetric matrices.
$\boldsymbol{R}$, $\boldsymbol{R}_0$, $\mathbfcal{R}$ and $\mathbfcal{R}_0$ are column vectors,
\begin{equation}
\boldsymbol{R} = \left(\begin{matrix} x \\ y \\ z \end{matrix}\right) =
R \left(\begin{matrix} \sin\theta\cos\phi \\ \sin\theta\sin\phi \\ \cos\theta
\end{matrix}\right)
\end{equation}
and similarly for $\boldsymbol{R}_0$, $\mathbfcal{R}$ and $\mathbfcal{R}_0$.
If the traps are anharmonic, the potentials for motion in $\boldsymbol{R}$ and $\mathbfcal{R}$ are also anharmonic. The coupling term is then more complicated, but is still zero if either $\boldsymbol{R}=\boldsymbol{R}_0$ or $\mathbfcal{R}=\mathbfcal{R}_0$. In the remainder of this paper, we neglect the coupling term and focus on the relative motion. This is a good approximation for the lowest trap state that is of principal interest for mergoassociation.

\section{Coupled-channel formulation for relative motion}
\label{sec:coupled-form}

\subsection{The trap potential}
\label{sec:relative}

\begin{figure}[tbp]
\begin{center}
\includegraphics[width=0.9\columnwidth]{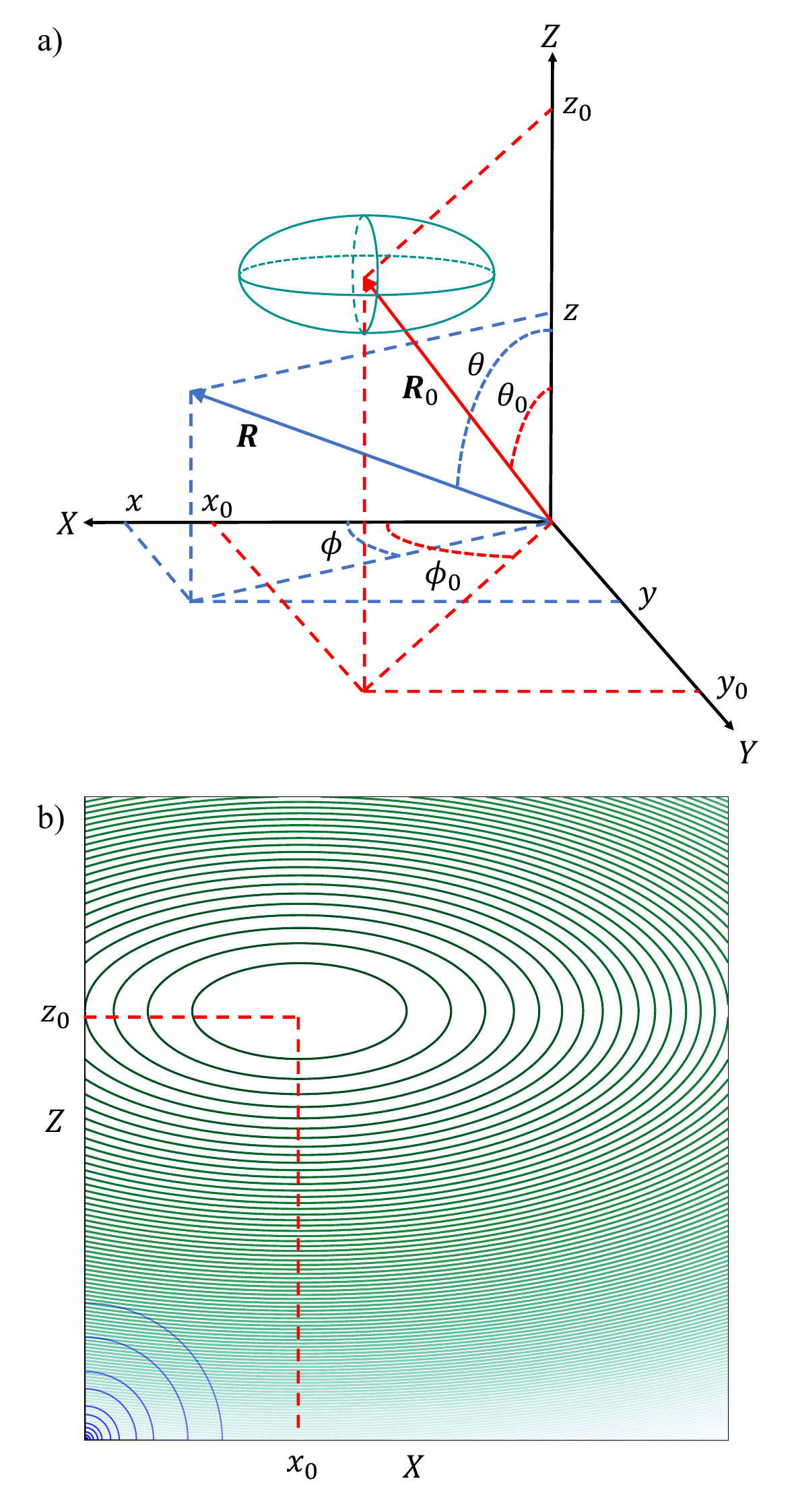}
	\caption{(a) Coordinate system for relative motion. The ellipsoid is a schematic representation of the shape of the trap potential for relative motion, and the Cartesian axes are aligned along its principal axes. (b) A cut through the potential for relative motion for $y=0$, showing the contours of the harmonic trap centered at $\boldsymbol{R}_0$ (green) and a shorter-range atom-atom potential centered at the origin (blue).
\label{fig:coord}
}
\end{center}
\end{figure}

Even if the two traps are not coaligned, the potential for relative motion is harmonic. It has three principal axes perpendicular to one another, which are used to define Cartesian axes $X,Y,Z$ as above. The resulting coordinate system is shown in Fig.\ \ref{fig:coord}(a).
The potential for relative motion may be written
\begin{equation}
	V_{\rm rel}^\textrm{trap}(\boldsymbol{R})=\textstyle{\frac{1}{2}} \mu [\boldsymbol{R}-\boldsymbol{R}_0]^T \boldsymbol{\omega}_\textrm{rel}^2 [\boldsymbol{R}-\boldsymbol{R}_0],
\label{eq:pot-Cart}
\end{equation}
with a minimum at the trap separation $\boldsymbol{R}=\boldsymbol{R}_0$. A cut through this is shown in green in Fig.\ \ref{fig:coord}(b).

For calculations in spherical polar coordinates, it is convenient to expand the potential for relative motion as
\begin{equation}
V_\textrm{rel}^\textrm{trap}(\boldsymbol{R}) = \sum_{\lambda\kappa} V_{\lambda\kappa}(R) C_{\lambda\kappa}(\theta,\phi),
\label{eq:Vexp-c}
\end{equation}
where
$C_{\lambda\kappa}(\theta,\phi) = [4\pi/(2\lambda+1)]^{1/2} Y_{\lambda\kappa}(\theta,\phi) $ are Racah-normalized spherical harmonics. For the potential (\ref{eq:pot-Cart}), the only non-zero terms in the expansion are
\begin{align}
V_{00}(R)&=\textstyle{\frac{1}{2}}\mu\bar{\omega}_\textrm{rel}^2 R^2
+\textstyle{\frac{1}{2}}\mu \boldsymbol{R}_0^T \boldsymbol{\omega}_\textrm{rel}^2 \boldsymbol{R}_0; \\
V_{10}(R)&=-\mu\omega_{\textrm{rel},z}^2 z_0 R; \\
V_{1{}\pm1}&=\mp \textstyle{\frac{1}{\sqrt{2}}} \mu (\omega_{\textrm{rel},x}^2 x_0  - i \omega_{\textrm{rel},y}^2 y_0) R; \\
V_{20}(R)&= \textstyle{\frac{1}{6}}\mu (2\omega_{\textrm{rel},z}^2-\omega_{\textrm{rel},x}^2-\omega_{\textrm{rel},y}^2) R^2; \\
V_{2\pm2}(R)&= \textstyle{\frac{1}{2\sqrt{6}}} \mu (\omega_{\textrm{rel},x}^2-\omega_{\textrm{rel},y}^2) R^2,
\end{align}
where
\begin{equation}
\bar{\omega}_\textrm{rel}^2 = \textstyle{\frac{1}{3}} \left(
\omega_{\textrm{rel},x}^2
+\omega_{\textrm{rel},y}^2
+\omega_{\textrm{rel},z}^2\right).
\end{equation}
The constant term in $V_{00}(R)$ involving $\boldsymbol{R}_0^T \boldsymbol{\omega}_\textrm{rel}^2 \boldsymbol{R}_0$ is chosen to place the minimum of the combined trap at zero energy. It is often convenient to express the trap potential in terms of harmonic lengths for relative motion. $\beta_{\textrm{rel},\alpha}=[\hbar/(\mu\omega_{\textrm{rel},\alpha})]^{1/2}$.

There are two special cases of the expansion that are of particular interest. If the traps are cylindrically symmetrical around the intertrap vector $\boldsymbol{R}_0$, $z$ may be chosen to lie along $\boldsymbol{R}_0$. Terms with $\kappa\ne0$ are then zero and the expansion may be replaced by a simpler one in terms of Legendre polynomials $P_{\lambda}(\cos\theta)$,
\begin{equation}
V_\textrm{rel}^\textrm{trap}(\boldsymbol{R}) = \sum_{\lambda} V_{\lambda0}(R) P_{\lambda}(\cos\theta).
\label{eq:Vexp-leg}
\end{equation}
If the individual traps are spherical, the term $V_{20}(R)$ is also zero. This is the case handled by Stock \emph{et al.}\ \cite{Stock:2003} and Ruttley \emph{et al.}\ \cite{Ruttley:2023}.

The expansion (\ref{eq:Vexp-c}) remains valid for anharmonic potentials, but in this case the expansion does not terminate and the coefficients $V_{\lambda\kappa}(R)$ must usually be evaluated by numerical quadrature.

\subsection{The interaction potential}
\label{sec:potential}

The interaction potential $V_\textrm{int}(\boldsymbol{R})$ between the two atoms may be represented at various levels of complexity. For unstructured atoms, it is isotropic, $V_\textrm{int}(R)$. When all the harmonic lengths $\beta_{\textrm{rel},\alpha}$ are large compared to the range of the potential, it may be sufficient to represent $V_\textrm{int}(\boldsymbol{R})$ as a point contact potential \cite{Huang:1957},
\begin{equation}
V_\textrm{int}(\boldsymbol{R}) = \frac{2\pi\hbar^2 a(E)}{\mu} \delta(\boldsymbol{R}) \frac{\partial}{\partial R} R,
\label{eq:contact}
\end{equation}
where the scattering length $a(E)$ may depend on energy if required. Such a contact potential may be implemented in coupled-channel calculations as a boundary condition on the log-derivative of the s-wave component of the wavefunction,
\begin{equation}
	\frac{d\psi_{00}}{dR} [\psi_{00}(R)]^{-1}=-1/a(E)
\end{equation}
at $R=0$. A contact potential affects only states with non-zero density at $R=0$, which here occurs only for states with a component in $M=0$.

More complicated treatments might include atoms or molecules with additional coordinates $\xi$ for internal structure, such as alkali-metal atoms including electron and nuclear spin and Zeeman effects. The interaction potential then depends on $\xi$ as well as $\boldsymbol{R}$ and $V_\textrm{int}(\boldsymbol{R},\xi)$ may itself be anisotropic. The total wavefunction would then be expanded in a basis set that includes functions for $\xi$, as for calculations on untrapped atom pairs \cite{Hutson:Cs2:2008}.

\subsection{Coupled-channel equations}
\label{sec:coupled-eq}

The Schr\"odinger equation for relative motion is
\begin{align}
\left[\frac{\hbar^2}{2\mu} \left(-R^{-1}\frac{d^2}{dR^2}R + \frac{\hat{L}^2}{R^2} \right) + V(\boldsymbol{R}) - E \right]
\Psi(R,\theta,\phi) =  0,
\label{eq:Schrod}
\end{align}
where $\hat{L}^2$ is the angular momentum operator for relative motion of the atoms and $E$ is
the total energy. The total potential energy is $V(\boldsymbol{R}) = V_\textrm{rel}^\textrm{trap}(\boldsymbol{R}) + V_\textrm{int}(\boldsymbol{R})$.
To solve Eq.\ \ref{eq:Schrod}, we expand the wavefunction as
\begin{equation}
\Psi(R,\theta,\phi) = R^{-1} \sum_{LM} \psi_{LM}(R) Y_{LM}(\theta,\phi),
\label{eq:psiexp}
\end{equation}
where $Y_{LM}(\theta,\phi)$ are spherical harmonics normalized to unity. Substituting the expansion (\ref{eq:psiexp}) into Eq.\ \ref{eq:Schrod} gives a set of coupled equations for the
channel functions $\psi_{LM}(R)$,
\begin{equation}\frac{d^2\psi_{LM}}{dR^2}
=\sum_{L'M'}\left[W_{LM,L'M'}(R)-{\cal E}\delta_{LL'}\delta_{MM'}\right]\psi_{L'M'}(R),
\label{eq:se-invlen}
\end{equation}
where $\delta_{ij}$ is the Kronecker delta, ${\cal E}=2\mu E/\hbar^2$ and
\begin{align}
&W_{LM,L'M'}(R) = \frac{L(L+1)}{R^2} \delta_{LL'} \delta_{MM'} \nonumber\\
&+ \frac{2\mu}{\hbar^2}\int_0^{2\pi} \int_0^\pi Y_{LM}^*(\theta,\phi)
V(R,\theta,\phi) Y_{L'M'}(\theta,\phi) \sin\theta \, d\theta \, d\phi.
\label{eqWij}
\end{align}
The contribution of $V_\textrm{rel}^\textrm{trap}(\boldsymbol{R})$ to $W_{LM,L'M'}(R)$ is
\begin{align}
\frac{2\mu}{\hbar^2} \sum_{\lambda\kappa}
V_{\lambda\kappa}(R) & (-1)^M [(2L+1)(2L'+1)]^{1/2}  \nonumber\\ \times &
\left(\begin{matrix} L & \lambda & L' \cr -M & \kappa & M' \end{matrix} \right)
\left(\begin{matrix} L & \lambda & L' \cr  0 &   0    & 0  \end{matrix} \right).
\label{eq:W}
\end{align}

If $\boldsymbol{R}_0$ lies along one of the principal axes of the traps, chosen as $Z$, the potential (\ref{eq:Vexp-c}) is symmetric with respect to a proper rotation $C_2(Z)$, so the quantity $(-1)^M$ is conserved and separate calculations may be performed for even and odd $M$. In addition, basis functions for $M\ne0$ are symmetrized,
\begin{equation}
\Phi_{LM}(\theta,\phi) = \textstyle{\frac{1}{\sqrt2}} \left[ Y_{LM}(\theta,\phi) \pm (-1)^M Y_{L{}-M}(\theta,\phi) \right]
\end{equation}
and separate calculations are carried out for + and $-$ symmetry. Only the functions of + symmetry for $M\ne0$ are coupled to those for $M=0$.
Parity is not conserved, so functions for both even and odd $L$ must be included.
If in addition $V_\textrm{rel}^\textrm{trap}(\boldsymbol{R})$ is cylindrically symmetric about the $z$ axis, the sum over $\kappa$ is limited to $\kappa=0$. The coupled equations are then diagonal in $M$.

\subsection{Solution of coupled equations}
\label{sec:solution}

We solve the coupled equations to find bound states using the package \textsc{bound} \cite{bound+field:2019,mbf-github:2022}. This propagates solutions of the coupled-channel equations for a trial energy from short range and from long range to a matching point $R_\textrm{match}$ in the classically allowed intermediate region. It then converges upon energies at which the wavefunction and its derivative are continuous at $R_\textrm{match}$, using the methods described in ref.\ \cite{Hutson:CPC:1994}.
The coupled equations are propagated from $R=0$ to $R_\textrm{match}\approx R_0$ using the fixed-step symplectic log-derivative propagator of Manolopoulos and Gray \cite{MG:symplectic:1995} with a step size of 25~\AA\ and from $R_\textrm{max}$ to $R_\textrm{match}$ using the variable-step Airy propagator of Alexander and Manolopoulos \cite{Alexander:1987}. The outer limit of integration is chosen as
\begin{align}
R_\textrm{max}
=\left[\sum_{\alpha=x,y,z} (\alpha_0+\rho\beta_{\textrm{rel},\alpha})^2\right]^{1/2},
\end{align}
where $\rho$ is typically 4.

The present coupled-channel approach differs from the treatment of Stock \emph{et al.}\ \cite{Stock:2003} in that it does not need basis sets for the interatomic distance $R$, which is handled efficiently by the propagation. The $R$-dependent coupling matrices in our formulation are much smaller than the Hamiltonian matrix in a basis set that includes functions for $R$.

The size of the spherical-harmonic basis set required depends on $R_0$ and the trap geometry, and is discussed below.

It would be straightforward to apply the coupled-channel method with a realistic atom-atom potential $V_\textrm{int}(R)$ in place of the contact potential. This would require a much smaller step size for the short-range part of the propagation, but the method would be otherwise unchanged.

\begin{figure*}
\begin{center}
\includegraphics[width=2.0\columnwidth]{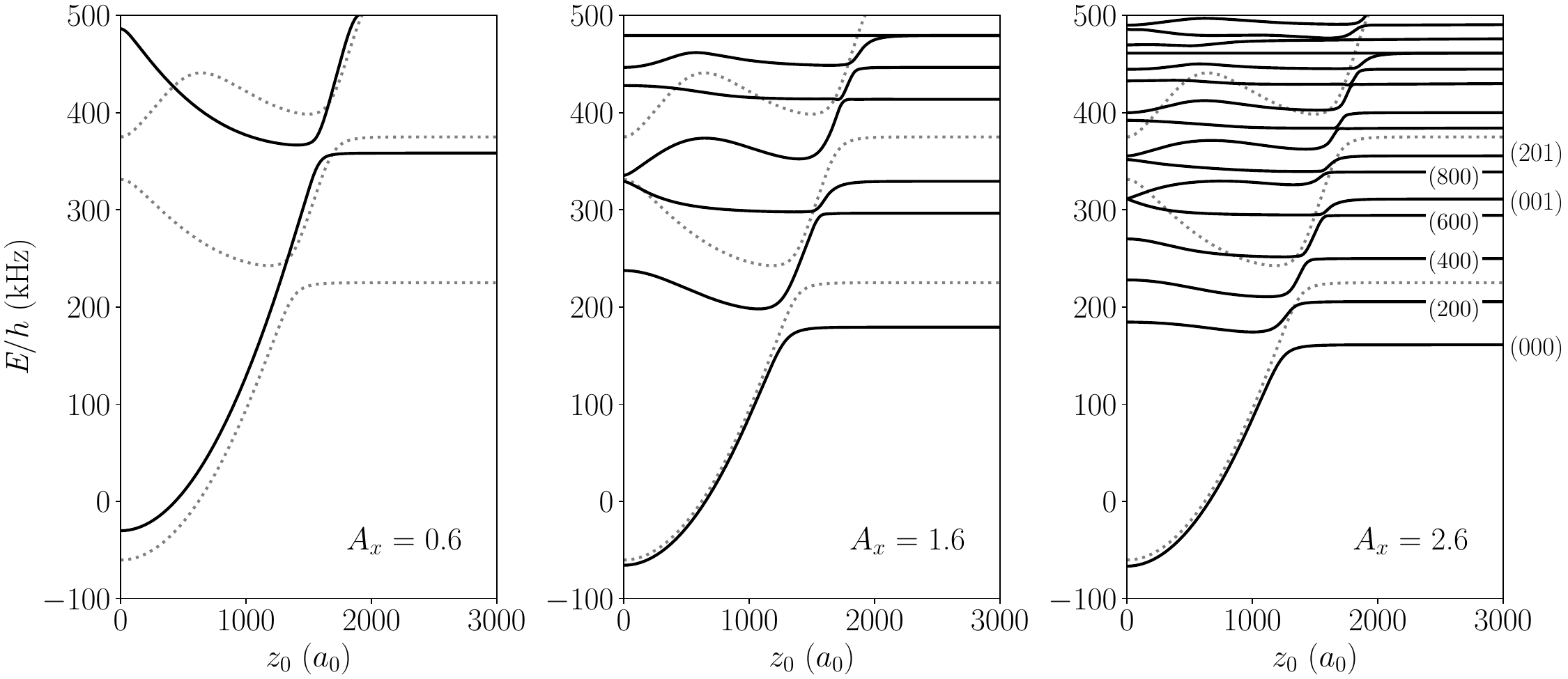}
\caption{Energies of two atoms in separate tweezers as a function of trap separation, obtained from coupled-channel calculations. Each panel is for a different aspect ratio $A_x$, for $\beta_{\textrm{rel},z} = \beta_{\textrm{rel},y} \approx 677\ a_0$, corresponding to $\omega_{\textrm{rel},z} = \omega_{\textrm{rel},y} = 150$~kHz. The grey dotted lines correspond to the spherical case with $A_x = A_y = 1$.
\label{fig:levels}
}
\end{center}
\end{figure*}

\section{Coupled-channel results}
\label{sec:coupled-res}

In this section we present coupled-channel results for a pair of cylindrically symmetric traps that approach one another along an axis perpendicular to their symmetry axis. This is close to the configuration that has been used experimentally to achieve mergoassociation of Rb and Cs atoms to form a weakly bound RbCs molecule \cite{Ruttley:2023}. It differs from the case considered in ref.\ \cite{Krych:2009}, where the traps approach along their symmetry axis. For consistency with \cite{Ruttley:2023}, we choose the axis $Z$ along the direction of approach and $X$ as the symmetry axis of the traps, with $\omega_{\textrm{rel},y}=\omega_{\textrm{rel},z}$. The separation of the traps is thus $z_0$, with $x_0=y_0=0$. We represent the atom-atom interaction with a contact potential of the form (\ref{eq:contact}), with a scattering length $a = 645\ a_0$ \cite{Takekoshi:RbCs:2012} appropriate for RbCs \footnote{This contact potential gives a binding energy $E_\textrm{b} = 83$~kHz for the least-bound state of RbCs; this somewhat underestimates the true value $E_\textrm{b} = 100\pm20$~kHz \cite{Takekoshi:RbCs:2012}, because the universal binding-energy formula $E_\textrm{b} = \hbar^2/(2\mu a^2)$ starts to break down at this depth \cite{Julienne:Li67:2014}.}. Since a contact potential affects only states with a component of $M=0$, we carry out calculations only for even $M$ and + symmetry.

We define aspect ratios $A_x = \beta_{\textrm{rel},x}/\beta_{\textrm{rel},z} = (\omega_{\textrm{rel},z}/\omega_{\textrm{rel},x})^{1/2}$ and $A_y = \beta_{\textrm{rel},y}/\beta_{\textrm{rel},z} = (\omega_{\textrm{rel},z}/\omega_{\textrm{rel},y})^{1/2}$. For the coupled-channel calculations in this section, with cylindrically symmetric traps, $A_y=1$.

The size of the basis set required depends on the trap geometry and also increases with $R_0$. For the majority of the calculations described here, including functions up to $L_\textrm{max}=24$ gives convergence of the energies to 6 significant figures for the largest $R_0$ considered here. Calculations for $A_x\ll 1$ required $L_\textrm{max}=40$.

Figure \ref{fig:levels} shows the energy levels for relative motion of two atoms in adjacent traps, as a function of trap separation, for various values of the aspect ratio $A_x$. In all cases, $\omega_{\textrm{rel},z}=150$ kHz. The dotted lines show the corresponding levels for a pair of spherical traps. At large separation, the energy levels of the trap states are those of a 3-dimensional harmonic oscillator in the relative motion. These are
\begin{align}
E&_{n_xn_yn_z} = \nonumber\\
&\quad\hbar[(n_x+\textstyle{\frac{1}{2}})\omega_{\textrm{rel},x} + (n_y+\textstyle{\frac{1}{2}})\omega_{\textrm{rel},y} +(n_z+\textstyle{\frac{1}{2}})\omega_{\textrm{rel},z}].
\label{eq:harm-3d-e}
\end{align}
The energies shown are those for states that feel the influence of the contact potential, which are those with non-zero density at $R=0$; for the trap states, this corresponds to limiting $n_x$ and $n_y$ to even values. The quantum numbers are shown for $A_x=2.6$ in Fig.\ \ref{fig:levels}; this corresponds to $\omega_{\textrm{rel},z} = 6.76\,\omega_{\textrm{rel},x}$, so trap levels with $(n_x,n_y,n_z)=(2,0,0)$, (4,0,0) and (6,0,0) all lie below (0,0,1) at large $z_0$. For small separation ($z_0 \lesssim \beta_{\textrm{rel},z}$), the trap states have substantial amplitude at $R=0$, so they are significantly shifted by $V_\textrm{int}(\boldsymbol{R})$. In the limit $z_0=0$, they correspond to the levels for two atoms in a cylindrically symmetric trap \cite{Bolda:2003, Idziaszek:2006}.

Cutting through the trap states is a molecular level that is shifted quadratically by the trap potential at $R=0$, which here is $\frac{1}{2}\mu\omega_{\textrm{rel},z}^2 z_0^2$. There is an additional shift due to the curvature of the trap potential, as described in Sec.\ \ref{sec:approx} below; this exists even at $z_0=0$. There are avoided crossings wherever the shifted molecular level would cross one of the trap levels.
It is the lowest of these avoided crossings that allows mergoassociation to form a weakly bound molecule from a pair of atoms; this occurs when traps containing atoms in their relative motional ground state are merged slowly enough to traverse the avoided crossing adiabatically.

The lowest crossing occurs at $z_0=z_0^\textrm{X}$, where the shifted molecular level has the same energy as the lowest level of the trap. When the atom-atom interaction is represented as a contact potential, this is approximately
\begin{equation}
z_0^\textrm{X} \approx \beta_{\textrm{rel},z} \left(1 + A_x^{-2} + A_y^{-2} + A_a^{-2}\right)^{1/2},
\label{eq:z_0_X}
\end{equation}
where $A_a=a/\beta_{\textrm{rel},z}$. We locate this crossing numerically using the state energies from coupled-channel calculations and then determine its precise position and effective coupling matrix element $\Omega_\textrm{eff}$ by a local fit of the energies near $z_0=z_0^\textrm{X}$ to the eigenvalues of a $2\times2$ matrix
\begin{equation}
\left(
\begin{matrix}
E_\textrm{X} + d_\textrm{mol}(z_0-z_0^\textrm{X}) & \Omega_\textrm{eff} \\
\Omega_\textrm{eff} & E_\textrm{X} + d_\textrm{at}(z_0-z_0^\textrm{X})
\end{matrix}
\right),
\end{equation}
where $E_\textrm{X}$ is the central energy of the avoided crossing and $d_{\rm at}$ and $d_{\rm mol}$ are the gradients of the atom-pair and molecular states near $z_0^{\rm X}$. To a first approximation, $d_\textrm{at}=0$ and
\begin{equation}
d_\textrm{mol}=\mu\omega_{\textrm{rel},z}^2 z_0^\textrm{X} = \frac{\hbar^2 z_0^\textrm{X}}{\mu\beta_{\textrm{rel},z}^4}.
\end{equation}
This procedure accurately determines the point of closest approach between the two states, and interprets their separation at that point as $2\Omega_\textrm{eff}$; however, it neglects effects due to other nearby states, so the resulting value of $\Omega_\textrm{eff}$ can be an underestimate of the true matrix element between the two states when other avoided crossings overlap the lowest one, as seen for $A_x=2.6$ in Fig.\ \ref{fig:levels}.

\begin{figure}[tbp]
\begin{center}
\includegraphics[width=0.9\columnwidth]{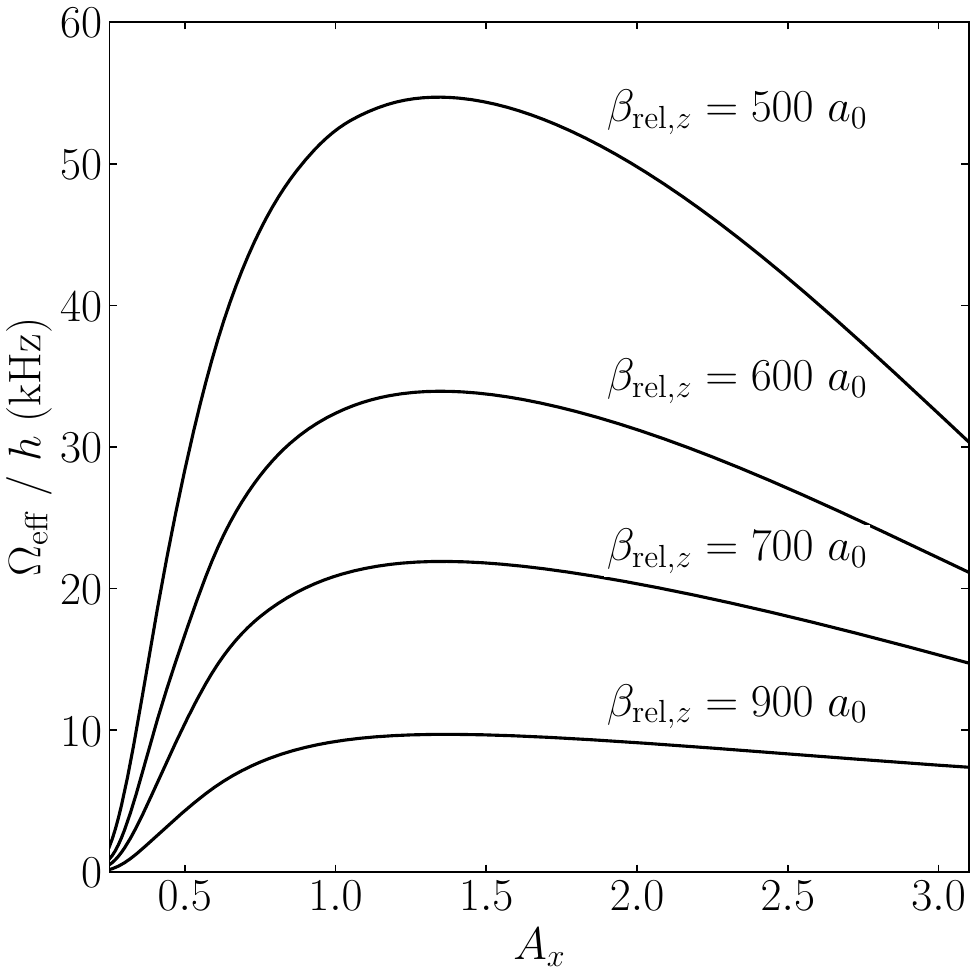}
\caption{Effective matrix element $\Omega_\textrm{eff}$ for the lowest avoided crossing, from coupled-channel calculations, as a function of aspect ratio $A_x$ for $A_y=1$ and various values of $\beta_{\textrm{rel},z}$.
\label{fig:Omega_eff}
}
\end{center}
\end{figure}

Figure \ref{fig:Omega_eff} shows the resulting values of $\Omega_\textrm{eff}$ as a function of aspect ratio $A_x$ for various values $\beta_{\textrm{rel},z}$.
The general form of $\Omega_\textrm{eff}$ for any value of $\beta_{\textrm{rel},z}$ is that it reaches a maximum at a value of $A_x$ near 1.3, corresponding to $\omega_{\textrm{rel},z}/\omega_{\textrm{rel},x} \approx 2$. There is a sharp dropoff in $\Omega_\textrm{eff}$ at smaller values of $A_x$, and $\Omega_\textrm{eff} \rightarrow 0$ as $A_x \rightarrow 0$. There is a much gentler dropoff at larger values of $A_x$. The origins of this behavior will be discussed in Section \ref{sec:approx}.

The semiclassical probability of traversing the avoided crossing adiabatically and thus forming a molecule may be calculated by numerical solution of the time-dependent Schr\"odinger equation. In a full treatment, this requires a derivative coupling matrix that may be obtained from the eigenstates as a function of $z_0$, which are available from the present calculation. A simple approximation to this is provided in the 2-state case by the Landau-Zener formula,
\begin{equation}
P_\textrm{LZ} = \exp{\left(\frac{-2\pi\Omega_\textrm{eff}^2}{\hbar \left|(d_\textrm{mol}-d_\textrm{at})\,dz_0/dt\right|}\right)},
\label{eq:LZ}
\end{equation}
where $dz_0/dt$ is the speed of relative motion of the traps at $z_0^\textrm{X}$. In initial experimental work, Ruttley \emph{et al.}\ \cite{Ruttley:2023} measured the probability of mergoassociation over the range $600\,a_0\lesssim \beta_{\textrm{rel},z} \lesssim 1500\,a_0$. They used tweezers with $A_x\approx2.6$, but nevertheless found that the probabilities were well reproduced using theoretical crossing strengths calculated for spherical traps ($A_x=1$). Figure \ref{fig:Omega_eff} explains this result: the strength of the avoided crossing dies off only slowly for $A_x>1.3$, and aspect ratios $1.5<A_x<2.6$ give crossing strengths qualitatively similar to those for $A_x=1$.

\section{Approximate model}
\label{sec:approx}

In this section we develop an approximate model that reproduces the main features of the coupled-channel results. The Hamiltonian for relative motion may be written
\begin{align}
\hat{H}_\textrm{rel}
&= \hat{T}_\textrm{rel} + V_\textrm{rel}^\textrm{trap}(\boldsymbol{R}) + V_\textrm{int}(\boldsymbol{R}) \nonumber\\
&= \hat{H}_\textrm{rel}^\textrm{trap} + V_\textrm{int}(\boldsymbol{R})
= \hat{H}_\textrm{int} + V_\textrm{rel}^\textrm{trap}(\boldsymbol{R}),
\end{align}
where $\hat{T}_\textrm{rel}$ is the kinetic energy operator, $\hat{H}_\textrm{rel}^\textrm{trap}$ is the Hamiltonian for the nonspherical harmonic trap and $\hat{H}_\textrm{int}$ is the Hamiltonian for the untrapped atom pair.
If $V_\textrm{int}(\boldsymbol{R})$ is represented as a contact potential as in Eq.\ \ref{eq:contact}, and $a>0$, $\hat{H}_\textrm{int}$ has a single molecular bound state, with eigenfunction
\begin{equation}
\psi_a = (2\pi a)^{-1/2} R^{-1} \exp(-R/a),
\label{eq:int-fun}
\end{equation}
and eigenvalue
\begin{equation}
E_a = -\hbar^2/(2\mu a^2).
\label{eq:int-e}
\end{equation}
The eigenfunctions of $\hat{H}_\textrm{rel}^\textrm{trap}$ are products of harmonic-oscillator functions in $x$, $y$ and $z$,
\begin{equation}
\psi_{n_xn_yn_z}(x,y,z) = \psi_{n_x}(x-x_0) \psi_{n_y}(y-y_0) \psi_{n_z}(z-z_0),
\label{eq:harm-3d-fun}
\end{equation}
where
\begin{align}
\psi_{n}(\alpha) = {}& (2^n n! \beta_{\textrm{rel},\alpha})^{-1/2} \pi^{-1/4} H_n(\alpha/\beta_{\textrm{rel},\alpha}) \nonumber\\
&\times\exp(-\textstyle{\frac{1}{2}} (\alpha/\beta_{\textrm{rel},\alpha})^2)
\label{eq:harm-1d-fun}
\end{align}
and $H_n(q)$ is a Hermite polynomial. The corresponding eigenvalues are given by Eq.\ \ref{eq:harm-3d-e}.

We consider a nonorthogonal basis set formed by the functions (\ref{eq:int-fun}) and (\ref{eq:harm-3d-fun}) and construct Hamiltonian and overlap matrices. The functions are normalized, so the diagonal elements of the overlap matrix ${\bf S}$ are all 1. The only non-zero off-diagonal elements are those between the bound-state function (\ref{eq:int-fun}) and the harmonic-oscillator functions (\ref{eq:harm-3d-fun}),
\begin{align}
S_{a,n_xn_yn_z} &= \langle a | n_xn_yn_z \rangle \nonumber\\
&=\int_0^{2\pi} \int_0^\pi \int_0^\infty \psi_a \psi_{n_xn_yn_z} r^2 dr\,\sin\theta d\theta \, d\phi.
\label{eq:S-int-harm}
\end{align}
These are evaluated by 3-dimensional numerical quadrature, using Gauss-Laguerre quadrature for $r$, Gauss-Legendre quadrature for $\theta$ and equally spaced and weighted points for $\phi$.

The diagonal elements of the Hamiltonian matrix for the harmonic-oscillator functions are
\begin{equation}
H_{n_xn_yn_z,n_xn_yn_z} = E_{n_xn_yn_z} + \langle n_xn_yn_z | V_\textrm{int}(\boldsymbol{R}) | n_xn_yn_z \rangle,
\end{equation}
where for a contact potential
\begin{align}
\langle n_xn_yn_z | & V_\textrm{int}(\boldsymbol{R}) | n_xn_yn_z \rangle = \nonumber\\
&(2\pi\hbar^2 a /\mu)
|\psi_{n_x}(x_0) \psi_{n_y}(y_0) \psi_{n_z}(z_0)|^2.
\end{align}
For the molecular function,
\begin{align}
H_{aa} = E_a + \langle a | V_\textrm{rel}^\textrm{trap}(\boldsymbol{R}) | a \rangle,
\end{align}
where
\begin{align}
\langle a | V_\textrm{rel}^\textrm{trap}(\boldsymbol{R}) | a \rangle =
V_\textrm{rel}^\textrm{trap}(\boldsymbol{R}_0)
+\frac{A_a^2}{12}\hbar\omega_{\textrm{rel},z} (1 + A_x^{-4} + A_y^{-4}). 
\label{eq:a_Vtrap_a}
\end{align}
The second term accounts for the curvature of the trap potential. It is usually relatively small for $A_a \lesssim 1$, but is independent of $\boldsymbol{R}_0$, and is responsible for the shift of the molecular state at $z_0=0$ seen in Fig.\ \ref{fig:levels}, particularly at $A_x=0.6$.

The off-diagonal elements of the Hamiltonian between harmonic-oscillator functions are
\begin{align}
H_{n_x'n_y'n_z',n_xn_yn_z} = (2\pi\hbar^2 a /\mu) &\psi_{n_x}(x_0) \psi_{n_y}(y_0) \psi_{n_z}(z_0) \nonumber\\
\times &\psi_{n_x'}(x_0) \psi_{n_y'}(y_0) \psi_{n_z'}(z_0),
\end{align}
while those between the harmonic-oscillator functions and the molecular function are
\begin{align}
H_{a,n_xn_yn_z} = {}& E_{n_xn_yn_z} S_{a,n_xn_yn_z} \nonumber\\
&-(\hbar^2/\mu) (2\pi/a)^{1/2}
\psi_{n_x}(x_0) \psi_{n_y}(y_0) \psi_{n_z}(z_0).
\label{eq:H-int-harm}
\end{align}

\begin{figure}
\begin{center}
\includegraphics[width=0.95\columnwidth]{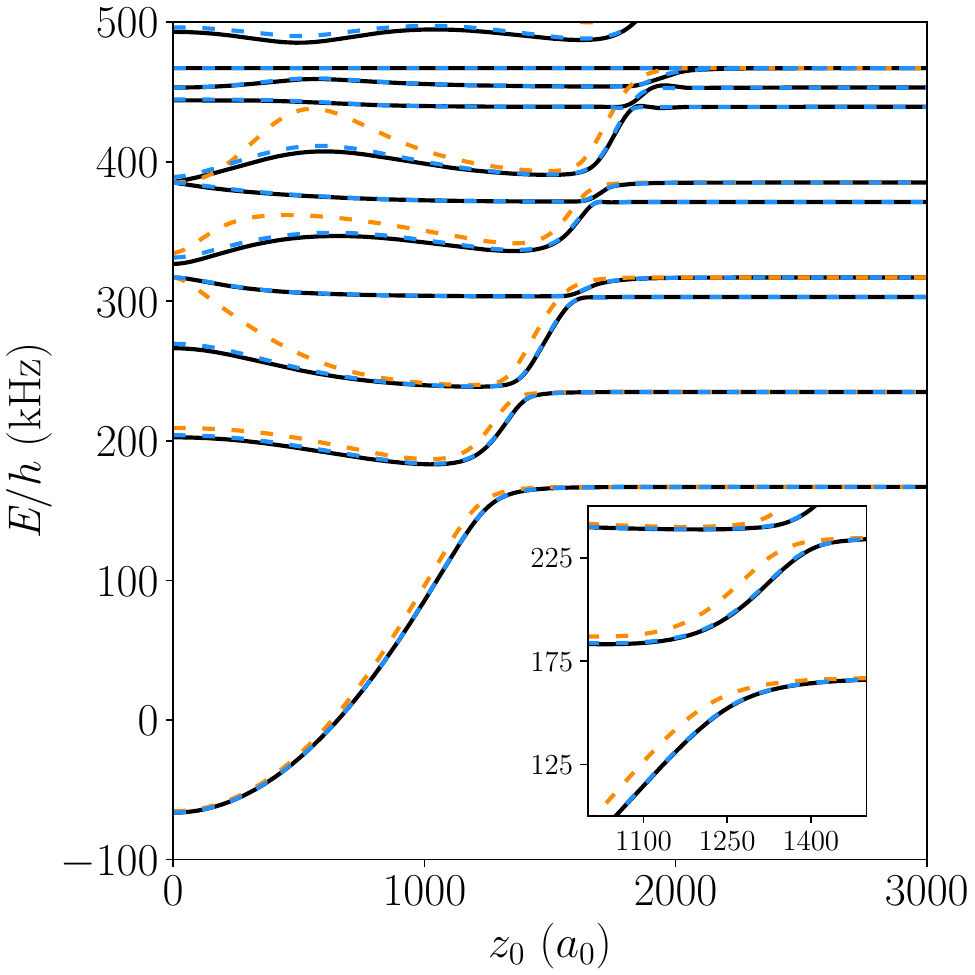}
\caption{Energies of two atoms in separate tweezers as a function of trap separation for $A_x=2.1$. Black lines show the results of coupled-channel calculations, while orange and blue dashed lines show basis-set calculations with $n_x,n_y,n_z \le 2$ and 10, respectively. Other parameters are as in Fig.\ \ref{fig:levels}.
\label{fig:levels-basis}
}
\end{center}
\end{figure}

The equations above may be used in two ways. First, they may be used to produce complete energy-level diagrams as a function of $R_0$ or other parameters. For this, matrices ${\bf H}$ and ${\bf S}$ are evaluated using a substantial number of harmonic-oscillator basis functions, and then used to solve a generalized matrix eigenvalue problem  ${\bf HC}=\bf{SCE}$ to produce eigenvectors ${\bf C}$ and a diagonal matrix of eigenvalues ${\bf E}$.  We illustrate this with the case investigated in Section \ref{sec:coupled-res}, with two cylindrically symmetric traps and the intertrap vector perpendicular to the symmetry axis of the traps. Figure \ref{fig:levels-basis} shows the levels for $A_x=2.1$, using harmonic-oscillator basis sets with $n_x,n_y,n_z \le 2$ and 10, compared with the results of coupled-channel calculations. It may be seen that the basis-set approach gives qualitatively correct results even for a small basis set. However, it is not fully converged for small $z_0$ even for a large basis set. This arises because the true wavefunctions have cusps at $\boldsymbol{R}=0$, due to the contact potential, and these cusps are poorly represented by an expansion in harmonic functions. They can be handled in spherical coordinates using parabolic cylinder functions in place of harmonic-oscillator functions \cite{Stock:2003}, but such functions are inefficient for well-separated traps.

\begin{figure}
\begin{center}
\includegraphics[width=0.9\columnwidth]{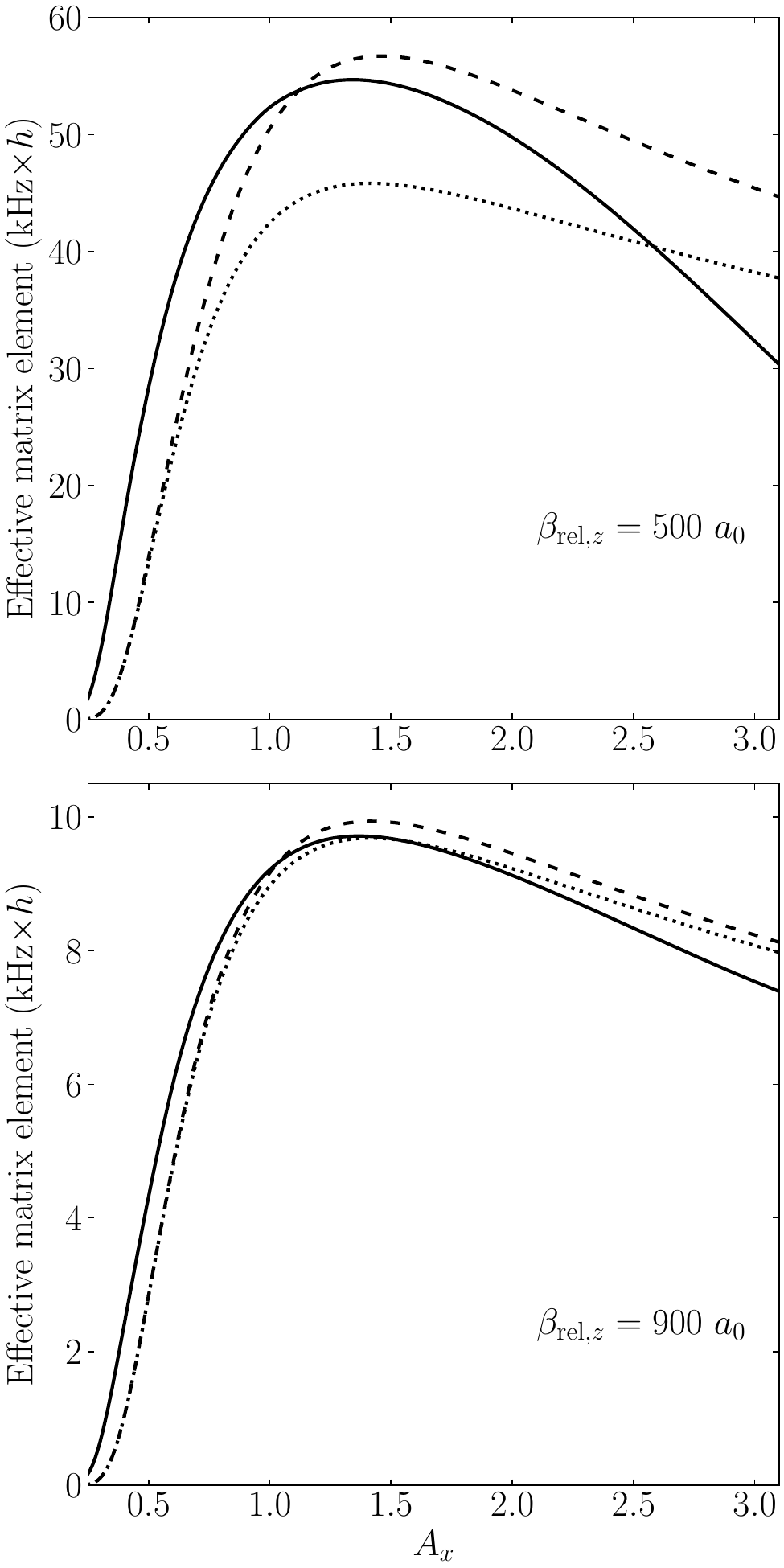}
\caption{Effective matrix element for the lowest avoided crossing as a function of aspect ratio $A_x$ for $A_y=1$ and two values of $\beta_{\textrm{rel},z}$; results shown are $\Omega_\textrm{eff}$ from coupled-channel calculations (black), $|\langle a | V_\textrm{int}(\boldsymbol{R}) | 000 \rangle|$ from Eq.\ \ref{eq:V-int-a-000-omega} (dotted), and $|\Omega_\textrm{eff}^{2\times2}|$ from Eq.\ \ref{eq:V_int_inc_S} (dashed).
\label{fig:cc-vs-approx}
}
\end{center}
\end{figure}

A much simpler application of the basis-set approach is to the strengths of avoided crossings that predominantly involve only the molecular state and a single harmonic-oscillator function. As in Sec.\ \ref{sec:coupled-res}, we focus on the crossing between the molecular state and the lowest harmonic-oscillator state. Under these circumstances, the off-diagonal matrix element of the Hamiltonian between the two functions is
\begin{align}
H_{a,000} = E_{000} S_{a,000} + \langle a | V_\textrm{int}(\boldsymbol{R}) | 000 \rangle,
\end{align}
where
\begin{align}
\langle a | V_\textrm{int}(\boldsymbol{R}) | 000 \rangle = -\frac{\hbar^2}{\mu}\left(\frac{2\exp[-\textstyle{\frac{1}{2}}(z_0^\textrm{X}/\beta_{\textrm{rel},z})^2]} {\sqrt{\pi}a\beta_{\textrm{rel},x} \beta_{\textrm{rel},y} \beta_{\textrm{rel},z}}\right)^{1/2}.
\label{eq:V-int-a-000}
\end{align}
If $z_0^\textrm{X}$ is taken from Eq.\ \ref{eq:z_0_X}, this may be written
\begin{align} -\hbar\omega_{\textrm{rel},z}\left(\frac{2\exp[-\textstyle{\frac{1}{2}}(1+A_x^{-2}+A_y^{-2}+A_a^{-2})]}
{\sqrt{\pi}A_x A_y A_a}\right)^{1/2}.
\label{eq:V-int-a-000-omega}
\end{align}
If the overlap integral $S_{a,000}$ is neglected, Eq.\ \ref{eq:V-int-a-000-omega} provides an analytic first approximation to the effective matrix element $\Omega_\textrm{eff}$, as shown by the dotted lines in Fig.\ \ref{fig:cc-vs-approx}. It also shows that the sharp dropoff in $\Omega_\textrm{eff}$ at small values of $A_x$ occurs because $z_0^\textrm{X}$ increases sharply as $A_x$ decreases, due to the term involving $A_x^{-2}$ in Eq.\ \ref{eq:z_0_X}. Conversely, the much slower dropoff at large $A_x$ occurs because of the harmonic-oscillator normalization factor involving $\beta_{\textrm{rel},x}^{-1/2}$ in Eq.\ \ref{eq:V-int-a-000}.

Equations \ref{eq:V-int-a-000} and \ref{eq:V-int-a-000-omega} are derived here for a contact potential. More generally, however, the wavefunction for a molecular state with binding energy $E_\textrm{b}$ is asymptotically of the form  $R^{-1}\exp(-kR)$, where $k=(2\mu E_\textrm{b}/\hbar^2)^{1/2}$ and $k^{-1}$ plays the role of an effective scattering length. This is valid even when $E_\textrm{b}$ is too large to be represented by Eq.\ \ref{eq:int-e} with the true scattering length $a$. For fixed $A_x$ and $A_y$, the quantity $\Omega_\textrm{eff}^2$ that appears in the Landau-Zener formula (\ref{eq:LZ}) is thus approximately proportional to
\begin{equation}
\omega_{\textrm{rel},z}^2 \tilde{E}_\textrm{b}^{1/2} \exp(-\tilde{E}_\textrm{b}),
\end{equation}
where $\tilde{E}_\textrm{b}=E_\textrm{b}/(\hbar\omega_{\textrm{rel},z})$. The strength of the avoided crossing decreases sharply for $\tilde{E}_\textrm{b}\gg1$, and the binding energies of the molecules that can be formed by mergoassociation are likely to be limited by the trap frequencies that can be achieved.

It may be noted that, for the case $A_y=1$ and neglecting overlap, Eq.\ \ref{eq:V-int-a-000-omega} predicts that the maximum value of $\Omega_\textrm{eff}$ appears at $A_x=\sqrt{2}$ for all values of $\beta_{\textrm{rel},z}$. This agrees remarkably well with the coupled-channel results in Fig.\ \ref{fig:Omega_eff}. The analytic expression shows a maximum for $A_x=\sqrt{2}$, qualitatively explaining the maximum near $A_x=1.3$ found from coupled-channel calculations.

If the effects of wavefunction overlap are included, the half-separation between the eigenvalues of the $2\times 2$ generalized eigenvalue problem at the point of closest approach is
\begin{equation}
\Omega_\textrm{eff}^{2\times2}=\frac{\langle a | V_\textrm{int}(\boldsymbol{R}) | 000 \rangle - \langle 000 | V_\textrm{int}(\boldsymbol{R}) | 000 \rangle S_{a,000}}{1-S_{a,000}^2}.
\label{eq:V_int_inc_S}
\end{equation}
This is nonanalytic, because the overlap integral $S_{a,000}$ must be evaluated by numerical quadrature. Nevertheless, the evaluation is straightforward. The values of $\Omega_\textrm{eff}^{2\times2}$ from Eq.\ \ref{eq:V_int_inc_S} are shown by the dashed lines in Fig.\ \ref{fig:cc-vs-approx}.

Figure \ref{fig:cc-vs-approx} shows that Eq.\ \ref{eq:V-int-a-000-omega} provides a qualitatively reasonable approximation to $\Omega_\textrm{eff}$ at large $\beta_{\textrm{rel},z}$, but that the approximation breaks down for smaller $\beta_{\textrm{rel},z}$, particularly for large $A_x$. Eq.\ \ref{eq:V_int_inc_S} improves the agreement when the overlap is moderate. However, both equations underestimate $\Omega_\textrm{eff}$ for $A_x<1$. This is due mainly to approximating $z_0^\textrm{X}$ by Eq.\ \ref{eq:z_0_X}, which neglects the second term in Eq.\ \ref{eq:a_Vtrap_a} and thus overestimates $z_0^\textrm{X}$. There are also remaining discrepancies at high $A_x$, particularly for smaller $\beta_{\textrm{rel},z}$. These arise because, for $z_0\lesssim 2\beta_{\textrm{rel},z}$ and $A_a \gtrsim 1$, the trap states are strongly shifted and mixed by $V_\textrm{int}(\boldsymbol{R})$. For $A_x^{-2}+A_a^{-2}\lesssim 1$, $z_0^\textrm{X}$ from Eq.\ \ref{eq:z_0_X} is small enough that this mixing is important and the lowest crossing is not well characterized by $H_{a,000}$ and $S_{a,000}$ alone. Under these circumstances it is necessary to use a larger basis set, rather than the $2\times2$ approximation implicit in Eqs.\ \ref{eq:V-int-a-000} and \ref{eq:V_int_inc_S}.

An important point to note is that, for a contact potential, the results may be expressed in dimensionless form, with all lengths (including $a$) expressed with respect to a single length scale ($\beta_{\textrm{rel},z}$ here) and all energies expressed with respect to a corresponding energy scale $\hbar\omega_{\textrm{rel},z}$. The results from both coupled-channel calculations and the basis-set approach are ``universal" for given values of $A_x$, $A_y$ and $A_a$ when expressed in these units. Results for values of $a$ that differ from $a=645\ a_0$ used here may thus be obtained by appropriate scalings of the harmonic lengths and energies, without additional calculations.

The coupled-channel approach of Sec.\ \ref{sec:coupled-form} can be applied for any interaction potential $V_\textrm{int}(\boldsymbol{R})$. However, the basis-set approach cannot be applied for interaction potentials that are non-integrable near $R=0$, as is the case for most realistic atom-atom potentials. It also cannot be applied for contact potentials corresponding to $a < 0$, because the molecular function (\ref{eq:int-fun}) then cannot be normalized. Furthermore, it requires very large basis sets of harmonic-oscillator functions when $A_a \gg 1$ and $R_0 \lesssim \beta_{\textrm{rel},z}$.

\section{Conclusions}
\label{sec:conc}

We have developed the theory of pairs of atoms in adjacent nonspherical traps. This is important for understanding mergoassociation \cite{Ruttley:2023}, in which weakly bound molecules are formed during the merging of two optical tweezers or cells of an optical lattice. For harmonic traps, we find that the separation of relative and center-of-mass motion is similar to that for two atoms in a single trap \cite{Deuretzbacher:2008}, but with a different coupling term between the motions.

We have developed a coupled-channel approach that can be used for the relative motion of atom pairs in harmonic traps with arbitrary anisotropy and arbitrary relative orientation. We have solved the coupled equations for pairs of coaligned nonspherical traps, as a function of trap separation. We approximate the atom-atom interaction here by a contact potential, but the method can be readily extended to handle other interaction potentials. If the molecule formed from the two atoms has a weakly bound state, it undergoes avoided crossings, as a function of trap separation, with the states of the trapped atom pair. Merging two traps that each contain an atom in its lowest motional state can thus form a molecule by adiabatic passage across the lowest-energy avoided crossing. This is mergoassociation.

We focus on the case important for mergoassociation with optical tweezers, where two traps that are individually cylindrical are merged along an axis $Z$ perpendicular to their symmetry axis $X$. The confinement along these axes is characterized by harmonic lengths $\beta_{\textrm{rel},z}$ and $\beta_{\textrm{rel},x}$, respectively, with aspect ratio $A_x=\beta_{\textrm{rel},x}/\beta_{\textrm{rel},z}$. The strength of the avoided crossing depends strongly on the aspect ratio: for fixed $\beta_{\textrm{rel},z}$, it has a maximum near $A_x=1.3$. In initial experimental work on mergoassociation \cite{Ruttley:2023}, it was found that experiments with $A_x\approx2.6$ were well reproduced by theory based on spherical traps ($A_x=1$). This is coincidental: $A_x=1$ and 2.6 give similar crossing strengths simply because they lie on opposite sides of the maximum.

We have developed an approximate model of the energy levels for separated traps. This uses a nonorthogonal basis set that combines a single molecular function with a set of Cartesian harmonic-oscillator functions for the trap states. The model gives reasonably accurate energy levels near the avoided crossing that is important for mergoassociation, though the harmonic-oscillator basis set converges slowly for small trap separations. In its simplest form, with only a single harmonic-oscillator function, the model gives an analytic expression for the crossing strength if overlap between the molecular and harmonic-oscillator functions is neglected. The analytic expression shows a maximum for $A_x=\sqrt{2}$, qualitatively explaining the maximum near $A_x=1.3$ found from coupled-channel calculations.

The methods developed in this paper will help understand and predict the efficiency of mergoassociation, both with optical tweezers and with transport in an optical lattice. This will allow efficient conversion of atom pairs into molecules for systems with weakly bound states, even if they do not possess resonances suitable for magnetoassociation. It may also be possible to extend mergoassociation to more complex systems, involving molecules or Rydberg atoms. The avoided crossing characterized here also offers opportunities for high-fidelity two-qubit quantum logic operations with atom pairs \cite{Jaksch:1999, Stock:2003, Stock:2006}.

\section*{Rights retention statement}

For the purpose of open access, the authors have applied a Creative Commons Attribution (CC BY) licence to any Author Accepted Manuscript version arising from this submission.

\section*{Data availability statement}

The data presented in this work are available from Durham University~\cite{DOI_data-merging-aniso}.

\section*{Acknowledgement}
We are grateful to Simon Cornish and Alex Guttridge for valuable discussions about the mergoassociation experiment.
This work was supported by the U.K. Engineering and Physical Sciences Research Council (EPSRC) Grant Nos.\ EP/P01058X/1, EP/T518001/1, EP/W00299X/1, and EP/V011677/1.

\bibliography{../all,merging-tweezers-data}
\end{document}